\documentclass[aps,prmaterials,superscriptaddress,twocolumn]{revtex4-2}
\usepackage{graphicx,amsmath,mathtools,amsfonts}
\usepackage{bm}
\usepackage[usenames, dvipsnames]{xcolor}
\usepackage{float}
\usepackage[english]{babel}
\usepackage{hyperref}
\hypersetup{colorlinks=true, linkcolor=blue, citecolor=blue, urlcolor=blue}

\begin{document}

\title{
Structural stability, electronic structure, and magnetism of the 
\\
$d^9$ double infinite-layer La$_3$Ni$_2$O$_5$F under chemical pressure and epitaxial strain
}

\author{K. Madani}
\author{Q. N. Meier}
\author{A. Cano}
\affiliation{Univ. Grenoble Alpes, CNRS, Grenoble INP, Institut N\'eel, 25 Rue des Martyrs, 38042, Grenoble, France}
\date{\today}

\begin{abstract}
Nickelate materials exhibit rich electronic properties that can be engineered toward cuprate-like regimes through topotactic and mixed-anion chemistry. Using first-principles calculations, we investigate the newly synthesized double infinite-layer oxyfluoride La$_3$Ni$_2$O$_5$F and its evolution under chemical pressure and epitaxial strain. 
The calculated phonon spectrum confirms the dynamical stability of the reported double infinite-layer crystal structure. Further, we find a highly two-dimensional cuprate-like Fermi surface dominated by Ni-$d_{x^2-y^2}$ states, with a moderate rare-earth-derived self-doping yielding an effective $\sim d^{1.2}_{x^2-y^2}$ filling. 
These electronic features remain remarkably robust under both chemical pressure and epitaxial strain. 
Spin-polarized calculations further reveal an extended manifold of nearly degenerate magnetic configurations with different in-plane and out-of-plane spin arrangements. 
Compressive strain further enhances this magnetic frustration while leaving the underlying electronic structure largely unchanged. Our results thus identify La$_3$Ni$_2$O$_5$F as a promising cuprate analogue and establish lattice engineering as an effective strategy for fine tuning its electronic and magnetic properties.
\end{abstract}

\maketitle

\section{Introduction}\vspace{-0.25cm}
The search for new high temperature superconductors analogous to the high-$T_c$ cuprates has recently led to the discovery of superconductivity in two distinct families of nickelates: the infinite-layer compounds~\cite{Li2019,Pan2022} and the multilayer Ruddlesden-Popper phases~\cite{Sun2023}.
As in the cuprates, the superconducting properties of these systems are difficult to reconcile with a conventional electron-phonon pairing mechanism~\cite{nomura2019, Meier2024,Ouyang2024,DeCataldo2024}. However, neither family realizes the cuprate analogy in a straightforward manner. In infinite-layer nickelates, rare-earth-derived conduction bands intersect the Fermi level, giving rise to self-doping of the nominally single Ni-$d_{x^2-y^2}$ band and thereby departing from the ideal $d^9$ electronic configuration~\cite{Lee2004,Botana2020}. In contrast, superconductivity in the reduced Ruddlesden-Popper nickelates emerges from an intrinsically multiband electronic structure, where both Ni-$d_{x^2-y^2}$ and Ni-$d_{z^2}$ states contribute to the low-energy physics \cite{Sun2023,Ouyang2024,deVaulx2025}.

\begin{figure}[t!]
\includegraphics[width=.27\textwidth]{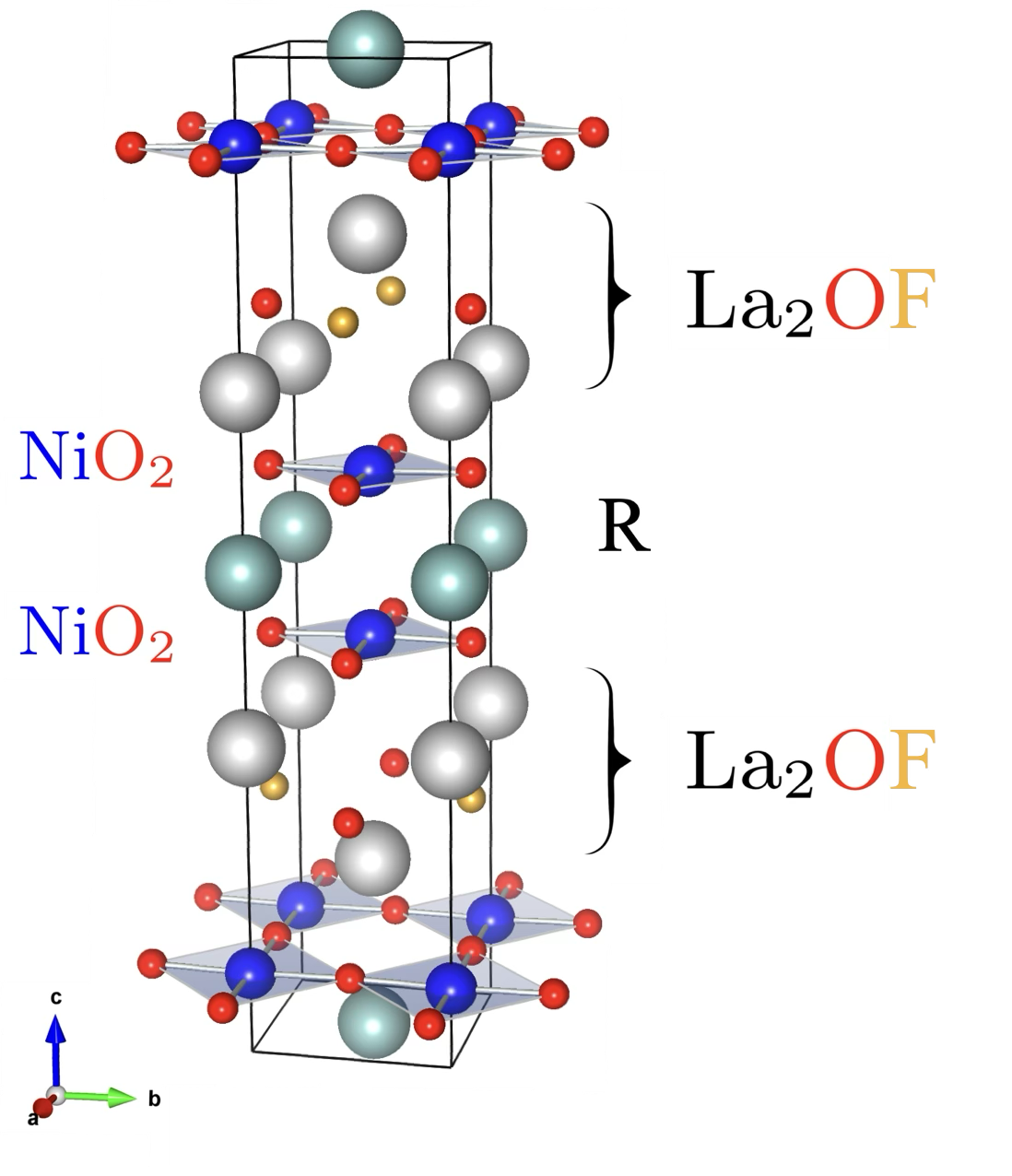}
\caption{Ball-and-stick model of the double infinite-layer nickelate La$_2R$Ni$_2$O$_5$F. 
The system comprises a bilayer NiO$_2$ block separated by fluorite-type La$_2$OF spacers. 
Within this block, the two square-planar NiO$_2$ layers are separated by rare-earth $R$ elements. Lines indicate the conventional cell for the O/F ordering considered throughout.}
\label{fig:structure}
\end{figure}

To realize a more genuinely cuprate-like electronic structure in nickelates, a mixed-anion strategy has recently emerged as a promising route. In particular, selective fluorination of Ruddlesden-Popper phases has been shown to suppress rare-earth-induced self-doping and to stabilize an essentially single-band, strongly two-dimensional Fermi surface, as demonstrated for the single-layer T' oxyfluoride La$_2$NiO$_3$F~\cite{Bernardini2021,Wissel2022}. Here, we extend this mixed-anion approach to the double infinite-layer counterpart La$_3$Ni$_2$O$_5$F (see Fig.~\ref{fig:structure}), recently synthesized via sequential topochemical reduction~\cite{Wernert2026}.
Using first-principles calculations, we characterize its electronic and magnetic properties and quantify the changes that can be obtained by means of chemical pressure and epitaxial strain to optimize its cuprate-like features.

\section{Methods}

We performed density-functional-theory (DFT) calculations using the Vienna Ab initio Simulation Package (\textsc{vasp})~\cite{kresse1996efficient,kresse1993ab,kresse1994ab,kresse1999ultrasoft,kresse1996efficiency,kresse1994norm}.
For the rare-earth elements other than La, the $4f$ electrons were treated as part of the frozen core. Specifically, the valence electronic configuration was $5p^6 6s^2 5d^1 $, with nine valence electrons treated explicitly in the projector augmented-wave (PAW) pseudo-potentials.

Non-magnetic calculations were performed using the PBE exchange-correlation functional~\cite{pbe_func}. We used the two-formula-unit conventional cell , containing a total of 22 atoms, an $800$~eV plane-wave cutoff, and a $\Gamma$-centered $14\times14\times4$ $k$-mesh, as determined by convergence tests.
Initial structures were taken from experimental data~\cite{Wernert2026} and then relaxed until Hellmann-Feynman forces were below $0.01$~eV/\AA\ (or $0.001$~eV/\AA\ for further phonon calculations). Epitaxial strain was simulated by fixing the in-plane lattice parameters $a=b$ to target values of $\varepsilon=(a-a_0)/a_0$, while fully relaxing $c$ and all internal coordinates.
For post-processing, \textsc{ifermi}~\cite{ifermi} was used for Fermi surface computation and \textsc{pymatgen}~\cite{pymatgen} for general structural and electronic-structure analysis.

Phonon spectra were obtained using the finite-displacement method as implemented in the \textsc{phonopy} package~\cite{phonopy1,phonopy2}, using atomic displacements of
$0.005$~\AA. Calculations were carried out mainly on a
$2(\sqrt{2}\times\sqrt{2})\times1$ supercell and cross-checked using a $2\times2\times2$ supercell, adjusting the $k$-point mesh density proportionally to match the enlarged real-space dimensions.

To evaluate the electronic response, we computed the bare electronic susceptibility $\chi_0 (\mathbf{q}) = \frac{2}{\Omega} \sum_{n',n,\mathbf{k}} \frac{
f(\varepsilon_{n'\mathbf{k}+\mathbf{q}}) - f(\varepsilon_{n\mathbf{k}})}{\varepsilon_{n'\mathbf{k}+\mathbf{q}} - \varepsilon_{n\mathbf{k}} - i\eta}$,
where $f$ is the Fermi-Dirac distribution and $\varepsilon_{n\mathbf{k}}$ the energy eigenvalues. By using this expression, we implicitly take the Bloch overlap matrix elements $\left| \langle \psi_{n\mathbf{k}} | e^{-i\mathbf{q}\cdot\mathbf{r}} | \psi_{n'\mathbf{k}+\mathbf{q}} \rangle \right|^2$ as unity~\cite{johannes_fermi-surface_2006, johannes_fermi_2008}.
This quantity is calculated within a dense $k$-point mesh grid and, to ensure numerical convergence, Fourier interpolation \cite{ifermi,MADSEN2018140} was applied to the energy eigenvalues across the Brillouin zone, yielding $\sim 25\times 10^{4}$ effective $k$-points.

Spin-polarized calculations were performed in a $\sqrt2\times\sqrt2\times1$ supercell to consistently compare the different magnetic configurations, using a $10\times10\times4$ $k$-mesh and an $800$~eV energy cutoff. In the case of LDA+$U$ and PBE+$U$, we used the Dudarev implementation~\cite{dudarev_electron-energy-loss_1998} with $U$ applied to the Ni $3d$ states.

\vspace{-.75em}

\section{Results}

\subsection{Structural stability}
We start by investigating the stability of the reported crystal structure.
La$_3$Ni$_2$O$_5$F crystallizes with a fluorite-type La$_2$OF spacer
separating double NiO$_2$ layers. 
Within the spacer, the O and F anions may in principle adopt different ordering patterns and induce structural distortions, as observed in the related mixed-anion compound La$_2$NiO$_3$F$_2$~\cite{Turkiewicz2026}.
We consider the particular O/F ordering
illustrated in Fig.~\ref{fig:structure}, which corresponds to the
highest-symmetry configuration compatible with the experimentally
determined lattice parameters, i.e., it preserves translational symmetry
with no superstructure.

\begin{figure}[t!]
\centering
\includegraphics[width=.95\linewidth]{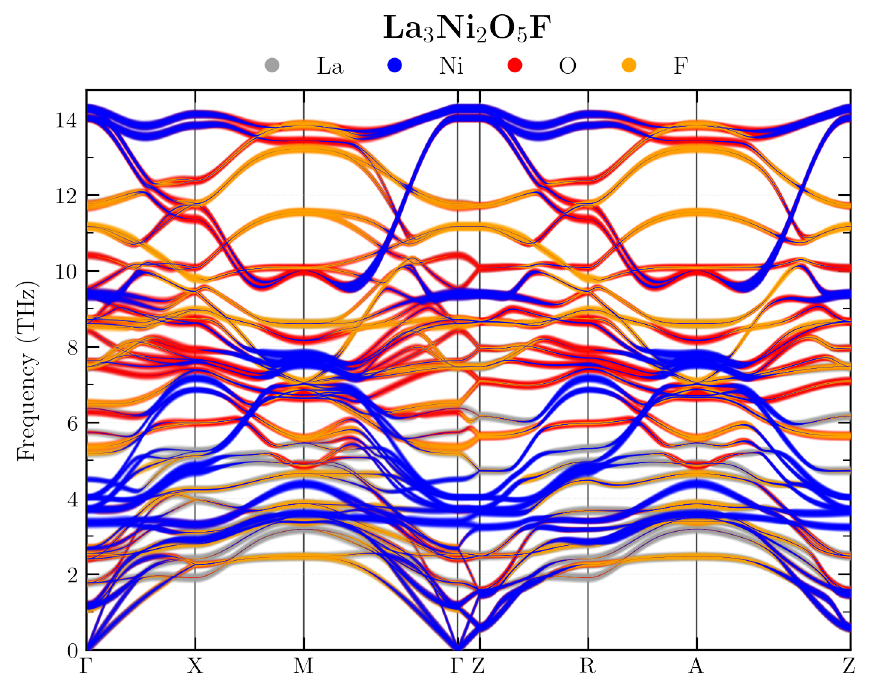}
\caption{Calculated phonon spectrum of La$_3$Ni$_2$O$_5$F. The colors indicate the main atoms involved in the corresponding modes.}
\label{fig:phonons}
\end{figure}

To assess the stability of this reference structure, we computed
the phonon spectrum using finite displacement method. The result, shown in Fig.~\ref{fig:phonons}, exhibits no imaginary frequencies across the entire Brillouin zone, confirming that the structure is stable. On this basis, we adopt the considered O/F order and optimized structure as the reference configuration for the remainder of this study, including the strain-dependent analysis.

\begin{figure*}[t]
\centering
\includegraphics[width=1\linewidth]{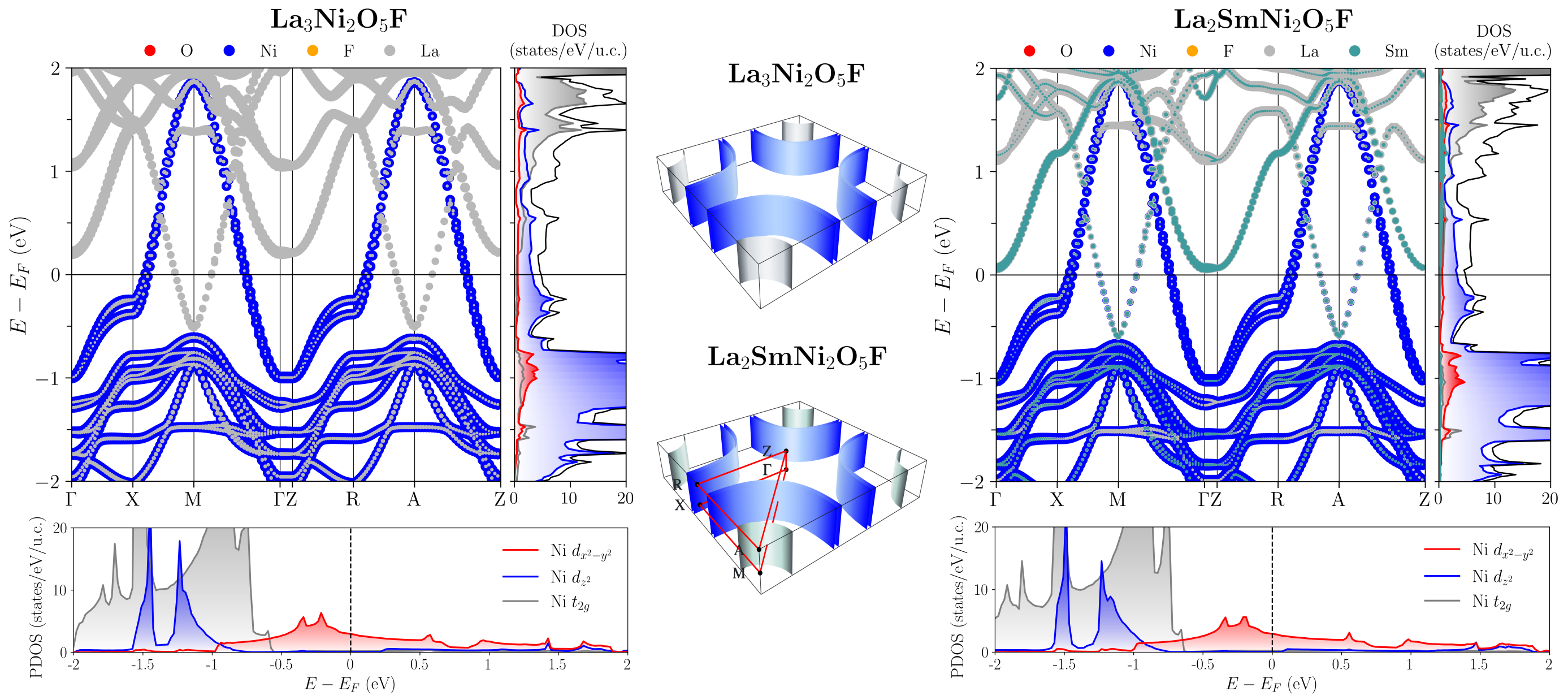}

\caption{Calculated electronic structure of (left) La$_3$Ni$_2$O$_5$F and (right) La$_2$SmNi$_2$O$_5$F, with the corresponding $k$-path shown as an inset. Orbital-resolved density of states (DOS) and Fermi surfaces are shown alongside the bands.}
\label{fig:electronic_bands}
\end{figure*}

\begin{table*}[t!]
\centering
\begin{tabular}{l cccc @{\hspace{0.8cm}} cccc @{\hspace{0.8cm}} cccc}
\hline\hline
 & \multicolumn{12}{c}{Occupancy per Ni} \\
\cline{2-13}
 & \multicolumn{4}{c}{Strain $-3\%$} & \multicolumn{4}{c}{Strain $0\%$} & \multicolumn{4}{c}{Strain $+3\%$} \\
& $d_{x^2-y^2}$ & $d_{z^2}$ & $t_{2g}$ & Total
& $d_{x^2-y^2}$ & $d_{z^2}$ & $t_{2g}$ & Total
& $d_{x^2-y^2}$ & $d_{z^2}$ & $t_{2g}$ & Total \\\hline
La$_3$Ni$_2$O$_5$F & 1.195 & 1.756 & 5.904 & 8.854 & 1.261 & 1.811 & 5.913 & 8.985 & 1.263 & 1.958 & 5.989 & 9.210 \\
La$_2$SmNi$_2$O$_5$F & 1.195 & 1.690 & 5.750 & 8.635 & 1.234 & 1.725 & 5.790 & 8.751 & 1.286 & 1.868 & 5.898 & 9.052 \\
\hline\hline
\end{tabular}
\caption{Orbital-resolved occupancy per Ni calculated for La$_3$Ni$_2$O$_5$F and La$_2$SmNi$_2$O$_5$F.}
\label{tab:occupations_strain}
\end{table*}

\subsection{Nonmagnetic electronic structure}
We now examine the nonmagnetic electronic structure of La$_3$Ni$_2$O$_5$F. Figure~\ref{fig:electronic_bands} (left panel) shows the calculated band structure, density of states, and Fermi surface. The low-energy states are dominated by Ni-$d_{x^2-y^2}$ character, with additional O-$2p$ and La-derived contributions. 
These shift the Ni occupation away from the nominal Ni$^{1+}$ ($d^9$) configuration, giving rise to a finite self-doping whose magnitude is summarized in Table~\ref{tab:occupations_strain}.

As in the single-layer $R_2$NiO$_3$F series~\cite{Wissel2022}, but unlike infinite-layer $R$NiO$_2$ nickelates~\cite{Botana2020,Bernardini2020}, the Ni-$d_{x^2-y^2}$ and $d_{z^2}$ manifolds remain well separated, with negligible mixing between these states near the Fermi level. Further, from the corresponding orbital energies, we obtain a charge-transfer energy $\Delta=\varepsilon_{d_{x^2-y^2}}-\varepsilon_{p_{x/y}}\simeq 3$~eV. This value is lower than in infinite-layer nickelates ($\sim4.4$~eV) and the single-layer La$_2$NiO$_3$F ($\sim3.6$~eV), bringing the double infinite-layer La$_3$Ni$_2$O$_5$F significantly closer to the cuprate regime ($\sim2.7$~eV)~\cite{Botana2020,Bernardini2021}.

The corresponding Fermi surface is markedly two-dimensional and consists of Ni-$d_{x^2-y^2}$ cylinders together with a La-$5d$-derived pocket around the $M$ point. This pronounced two-dimensionality originates from the La$_2$OF fluorite spacers (comprising the strongly electronegative F ions), which suppress hybridization along the $c$ axis, as previously found for La$_2$NiO$_3$F~\cite{Bernardini2021}. 
In La$_3$Ni$_2$O$_5$F, in addition, the interlayer coupling remains weak as the two Ni-$d_{x^2-y^2}$ bands are nearly degenerate, with no clear bonding-antibonding splitting. 
This behavior contrasts with both the bilayer La$_3$Ni$_2$O$_7$ precursor~\cite{Sun2023,Sakakibara2023,deVaulx2025} and infinite-layer heterostructures~\cite{Bernardini2020h}, where stronger interlayer hybridization produces sizable splittings.

The La-$5d$-derived pocket is likewise highly two-dimensional. Although derived primarily from La-$5d$ orbitals, its charge density is concentrated in the interstitial region between adjacent NiO$_2$ planes rather than on the La sites themselves (Fig.~\ref{fig:Psi_M}). 
This pocket produces a self-doping effect that is intermediate between the nearly undoped La$_2$NiO$_3$F~\cite{Bernardini2021,Wissel2022} and the infinite-layer nickelates, where two rare-earth-derived pockets whose size is further amplified by electronic correlations contribute to the Fermi surface~\cite{Olevano2020}.

\begin{figure*}[t!]
\centering
\includegraphics[width=.975\linewidth]{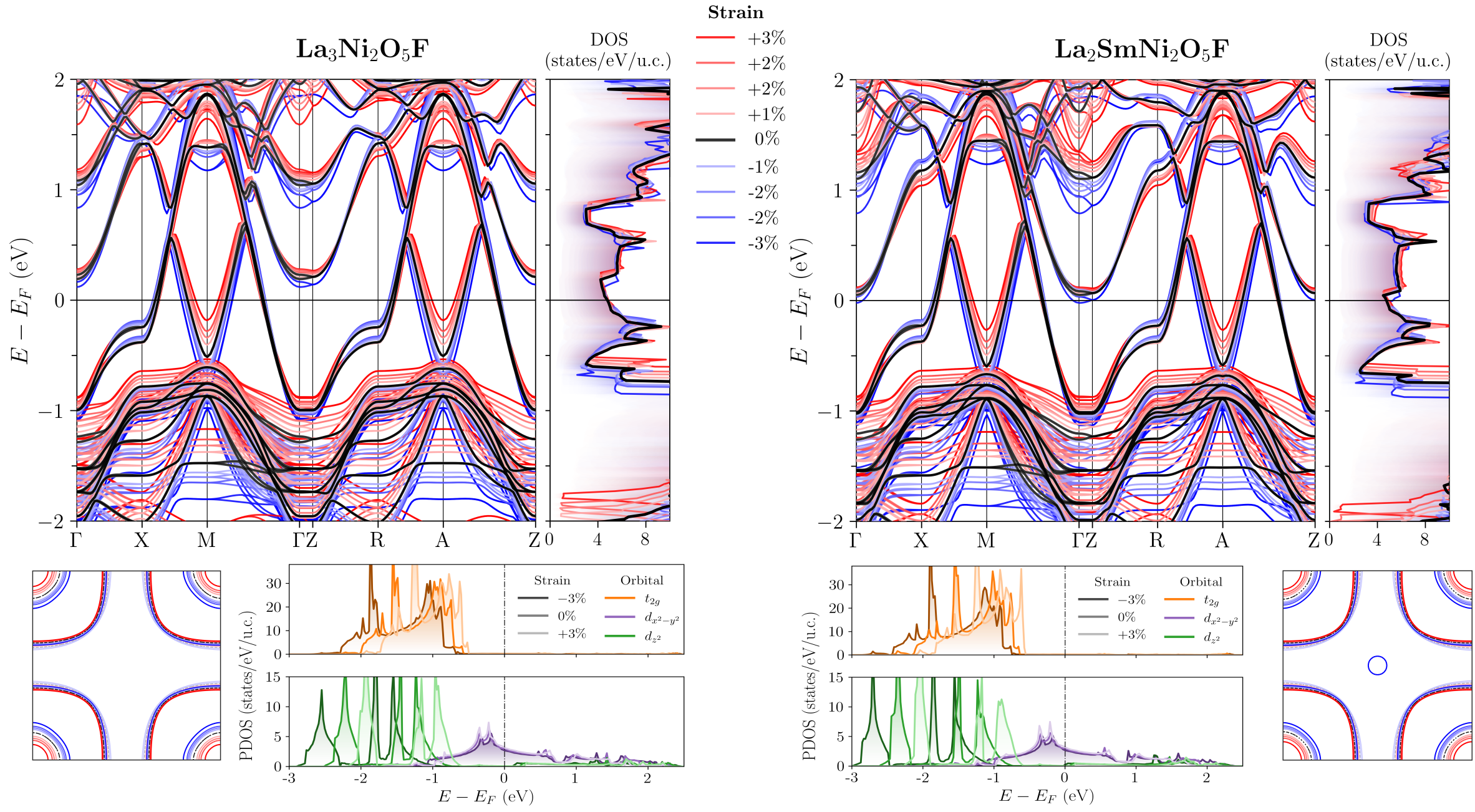}
\caption{Electronic band structures, orbital-resolved Ni DOS, and Fermi
surfaces calculated for La$_3$Ni$_2$O$_5$F (left) and
La$_2$SmNi$_2$O$_5$F (right) as a function of epitaxial strain.}
\label{fig:electronic_bands_strain}
\end{figure*}

\subsection{Chemical pressure}

To assess the robustness of these electronic features, we consider the substitution of La with smaller rare-earth ions. The resulting chemical pressure, driven by steric contraction~\cite{bernardini2021g}, is expected to modify the crystal structure and, consequently, the electronic properties. Such substitutions have also been shown experimentally to help select the desired Ruddlesden-Popper polymorph precursor, avoiding intergrowth issues~\cite{li_ambient_nodate}. The calculated lattice parameters confirm this chemical-pressure effect (Fig.~\ref{fig:lattice_params}). In the following we focus on La~$\to$~Sm substitution, for which $a$ and $c$ contract by 0.75\% and 1.5\%, respectively (see Fig.~\ref{fig:lattice_params} for other La~$\to R$ substitutions).

Figure~\ref{fig:electronic_bands} shows the calculated electronic structure of La$_2$SmNi$_2$O$_5$F. Compared with La$_3$Ni$_2$O$_5$F, the Ni-$d_{x^2-y^2}$ bands and the O-$2p$ manifold remain essentially unchanged, while the rare-earth-derived bands shift subtly downward. Specifically, the self-doping band sinks further below $E_F$ around $M$ and $A$, and the empty rare-earth conduction band moves closer to $E_F$ along $\Gamma$-$Z$. Although the absolute position of these states might be sensitive to the $4f$ treatment of the rare-earth ions, their evolution with chemical pressure is robust, yielding a consistent trend across the rare-earth series (Fig.~\ref{fig:bands_Re}).
Otherwise, despite the reduced NiO$_2$ interlayer spacing, the resulting Fermi surface closely resembles that of the La parent compound and retains its two-dimensional character, with only modest changes in the Ni-$d_{x^2-y^2}$ filling (Table~\ref{tab:occupations_strain}).

\subsection{Epitaxial strain}

Epitaxial strain provides a complementary route to tune the lattice parameters (Fig.~\ref{fig:lattice_params}) and thereby the electronic structure. The corresponding band structures of La$_3$Ni$_2$O$_5$F and La$_2$SmNi$_2$O$_5$F are shown in Fig.~\ref{fig:electronic_bands_strain}. In both compounds, the Ni-$d_{x^2-y^2}$ band remains remarkably robust under strain, preserving the cuprate-like low-energy electronic structure. The rare-earth-derived conduction bands, in contrast, are substantially more strain-sensitive, making epitaxial strain an effective means of controlling the self-doping.
Specifically, tensile strain ($\varepsilon > 0$) shifts the rare-earth bands upward, reducing its contribution at $E_F$, whereas compressive strain ($\varepsilon < 0$) shifts them downward and enhances the self-doping. The total filling of the Ni-$3d$ derived states decreases slightly for compressive strain and increases for tensile one. Interestingly, the increased filling under tensile strain is predominantly absorbed by the $d_{z^2}$ orbital and not the $d_{x^2-y^2}$, while compressive strain predominantly hole-dopes the $d_{x^2-y^2}$ orbital (see Fig.~\ref{fig:electronic_bands} and Table~\ref{tab:occupations_strain}). Otherwise, epitaxial strain primarily modifies the rare-earth pockets at M, while leaving the Ni-$d_{x^2-y^2}$ cuprate-like Fermi surface largely unaffected (see Fig. \ref{fig:electronic_bands_strain}).

For large compressive strains, the rare-earth $5d$ bands form an additional three-dimensional pocket at $\Gamma$, reminiscent of the rare earth pocket in the infinite layer nickelates \cite{Lee2004,Botana2020}. This pocket is expected to evolve into a cylinder upon further increased strain.

\begin{figure*}[t]
\centering
\includegraphics[width=0.73\linewidth]{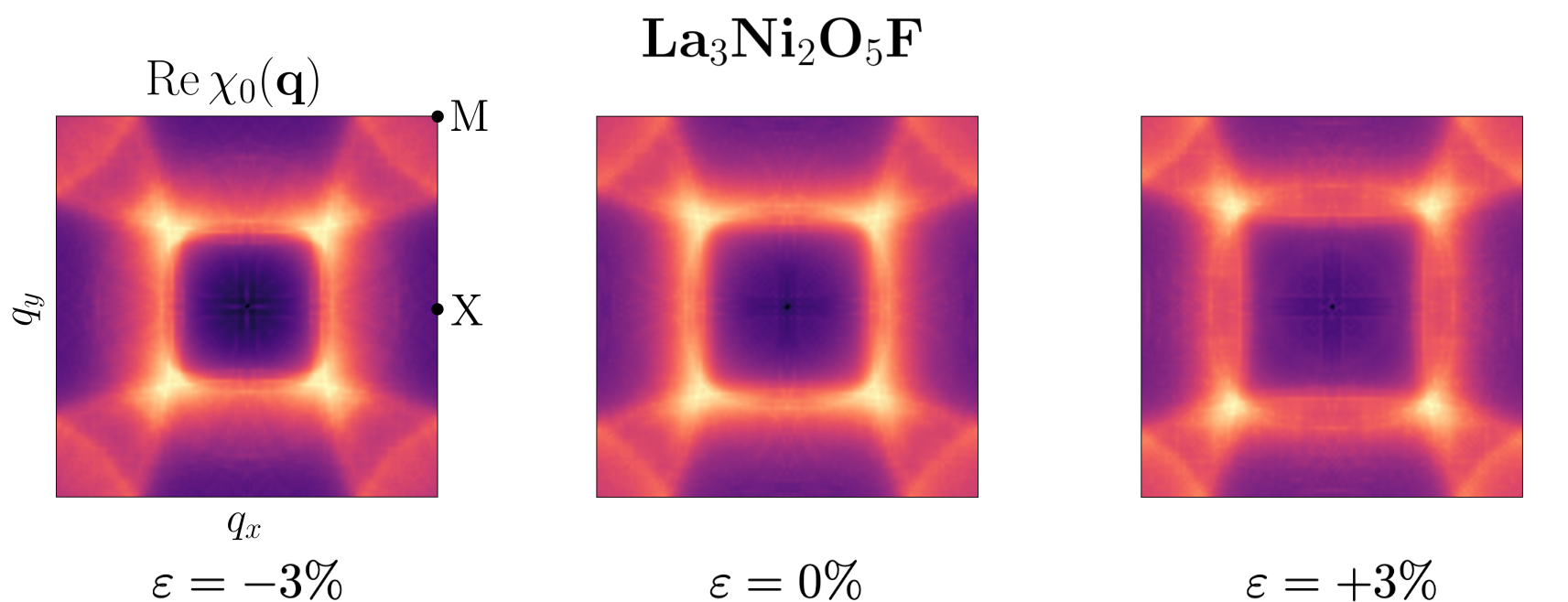}
\caption{Real part of the static
electronic susceptibility $\chi_0(\mathbf{q})$ of La$_3$Ni$_2$O$_5$F calculated for different values of epitaxial strain, $\varepsilon=-3$~\% (left), $0$~\% (middle), and $-3$~\% (right). 
}
\label{fig:electronic_instabilities}
\end{figure*}

\subsection{Electronic instabilities}

The La$_3$Ni$_2$O$_5$F Fermi surface displays intriguing cuprate-like features, including its nesting properties. This motivates a closer examination of the associated electronic instabilities and their strain dependence. 

Figure~\ref{fig:electronic_instabilities} shows the calculated static susceptibility $\chi_0(\mathbf{q})$ in the $q_z=0$ plane as a function of epitaxial strain. At $\varepsilon=0$, $\chi_0(\mathbf{q})$ displays pronounced peaks along the $\Gamma$-M directions at $\mathbf{q}_\text{peak}\simeq(0.23,0.23,0)$ (in units of $2\pi/a$). This is accompanied by a broader features parallel to $\Gamma$-X. 
These peaks shift with strain, moving inward under compression ($\varepsilon <0$) and outward under tension ($\varepsilon >0$) such that $\mathbf{q}_\text{peak}(\varepsilon)\simeq \mathbf{q}_\text{peak}(0)+ (\varepsilon,\varepsilon,0)$. 
This trend is consistent with the subtle strain evolution of the Fermi surface itself (Fig.~\ref{fig:electronic_bands_strain}), though it is more apparent in the nesting function (Fig.~\ref{fig:dd-app}): the relevant sheets contract under compressive strain, yielding a smaller nesting wavevector, and expand under tensile strain, yielding a larger one. Together, these results indicate that the underlying nesting condition, and hence the wavevector of a potential electronic instability, is continuously tunable by epitaxial strain. 

To assess whether this enhanced susceptibility can drive a global instability, we analyze in more detail the corresponding phonon spectra (Figs.~\ref{fig:phonons} and \ref{fig:phonons-app}). 
The phonon frequencies display the expected overall trend, with softening under tensile strain and hardening under compressive strain. However, no imaginary frequencies are observed, including under epitaxial strain. Moreover, we do not find any Kohn-anomaly-like feature along either the $\Gamma$-M or $\Gamma$-X directions. These results indicate that, although the electronic susceptibility is locally enhanced, it remains insufficient to induce an electronically-driven structural instability La$_3$Ni$_2$O$_5$F.

\subsection{Magnetism}
\begin{figure*}[t]
    \centering
    \includegraphics[width=\linewidth]{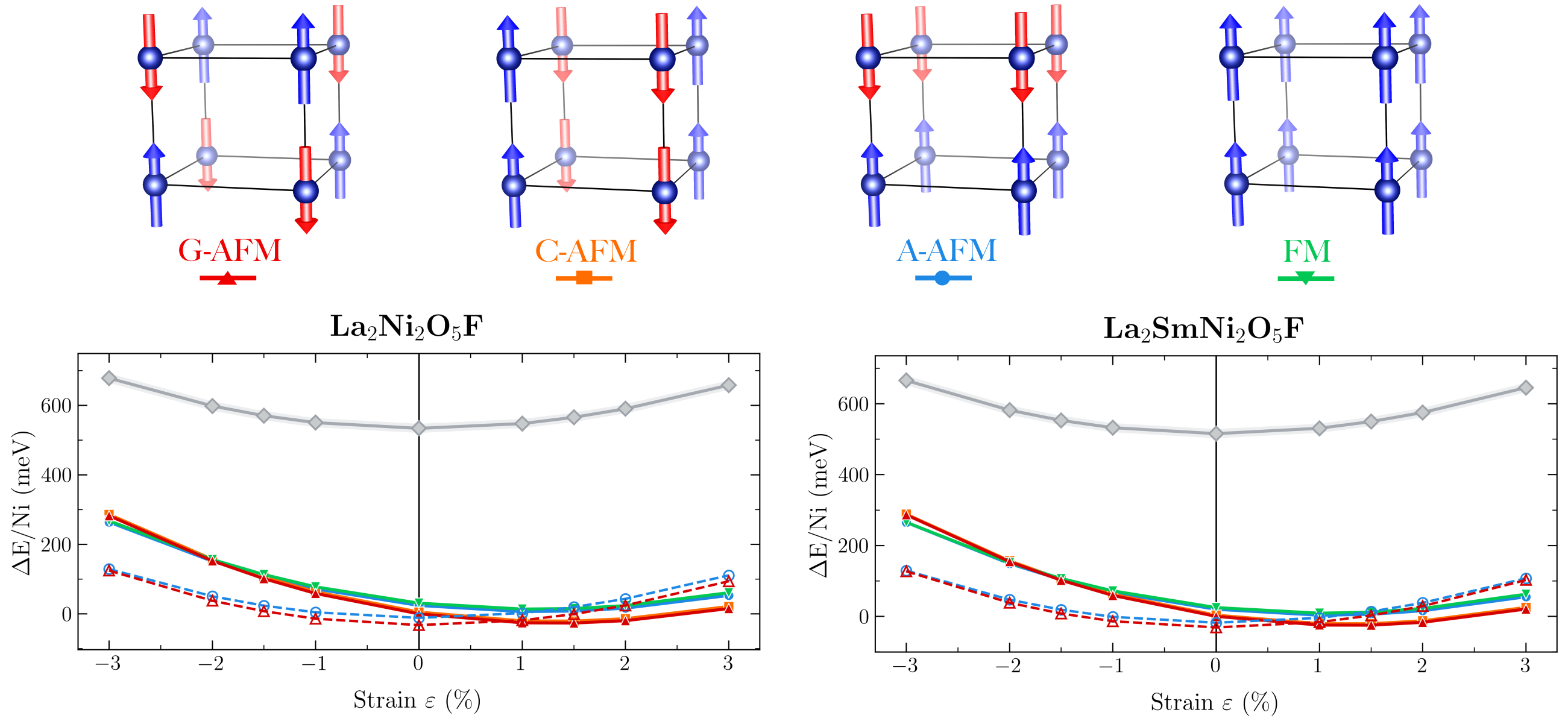}
    \caption{
Magnetization energies as a function of epitaxial strain for La$_3$Ni$_2$O$_5$F (left) and La$_2$SmNi$_2$O$_5$F (right), calculated within PBE+$U$ ($U=3.5$~eV). The non-spin-polarized calculation is shown in gray. Two sets of calculations are compared: spin-polarized calculations using the structure relaxed in the nonmagnetic state (solid symbols and lines), and calculations in which the structure is fully relaxed for each magnetic configuration (open symbols and dashed lines). The resulting shift of the energy minima by $\varepsilon \simeq 1.5 \%$ reveals a sizable magnetostructural coupling. Beyond this coupling, the calculations reveal a quasi-degeneracy among competing magnetic configurations that can be further tuned by epitaxial strain.
}.
    \label{fig:magnetism}
\end{figure*}

Finally, we investigate the possibility of magnetic ordering in La$_3$Ni$_2$O$_5$F and its evolution under chemical pressure (La $\rightarrow$ Sm substitution) and epitaxial strain. Specifically, we compare the nonmagnetic state with ferromagnetic (FM) and A-, C-, and G-type antiferromagnetic (AFM) configurations.

Figure~\ref{fig:mag_LDA_GGA_appendix} summarizes the magnetic energetics of La$_3$Ni$_2$O$_5$F as a function of the exchange-correlation functional. Within LDA, none of the collinear magnetic configurations is stabilized with respect to the nonmagnetic state. PBE instead favors magnetic order. Specifically, the G- and C-AFM configurations develop Ni moments of $\sim0.68\,\mu_B$, while in the A-AFM and FM ones these moments remain much smaller, $\sim0.23\,\mu_B$. 
The G-AFM configuration is the ground state, only $79$~meV/Ni below the nonmagnetic solution and nearly degenerate with C-AFM. A-AFM and FM are likewise mutually degenerate but closer in energy to the nonmagnetic state. Since each pair differs only in interlayer spin alignment, the magnetic energetics are governed predominantly by in-plane exchange within the NiO$_2$ block, with only weak coupling along $c$.

This near-degeneracy persists, and is even enhanced by correlations (Fig.~\ref{fig:mag_LDA_GGA_appendix} and \ref{fig:magnetism}). 
The ordered moments naturally increase in both LDA+$U$ and PBE+$U$, reaching $1\,\mu_B$ for $U=3.5$~eV. However, the energy differences among the competing states strikingly decrease. 

Chemical pressure (La~$\to$~Sm) leaves this picture essentially unchanged. 
However, compressive epitaxial strain further suppresses the residual magnetic energy scale, driving the system toward an almost fully degenerate manifold of collinear configurations without significantly changing the local moments (Figs. \ref{fig:magnetism} and \ref{fig:magnetism_appendix}).

Interestingly, full relaxation of the lattice parameters and internal
coordinates for the magnetic configurations reveals a sizable
magnetostructural coupling.
Without this relaxation (solid symbols in Fig.~\ref{fig:magnetism}), the energy minima of the different magnetic configurations are at $\varepsilon \approx 1.5~\%$ (since $\varepsilon=0$ is defined with respect to the nonmagnetic reference structure in that case). 
This offset reflects the magnetostructural coupling associated with the change in the in-plane lattice parameter upon magnetic ordering.
However, despite this sizable magnetostructural response, the relative magnetization energies and the resulting magnetic frustration remain essentially unchanged.

\section{Discussion}

Our results place La$_3$Ni$_2$O$_5$F within the emerging family of mixed-anion nickelates, bridging the single-layer T$'$ oxyfluorides $R_2$NiO$_3$F and the superconducting infinite-layer nickelates. 
Relative to the former, the bilayer compound preserves the strongly two-dimensional Ni-$d_{x^2-y^2}$ electronic structure while introducing a moderate self-doping contribution from rare-earth-derived states. 
Relative to the latter, however, this self-doping remains substantially weaker, leaving the electronic structure closer to the cuprate limit. The reduced charge-transfer energy, together with the clean separation of the $e_g$ manifold, further reinforces this trend, as both quantities have been associated with enhanced superconductivity in the cuprates~\cite{Aoki2012}.

Moreover, despite its bilayer structure, La$_3$Ni$_2$O$_5$F exhibits essentially no bonding-antibonding splitting of the Ni-$d_{x^2-y^2}$ bands. This is in sharp contrast to infinite-layer heterostructures \cite{Bernardini2020h} and its bilayer precursor La$_3$Ni$_2$O$_7$, where such splitting underpins the proposed $s_\pm$ pairing scenario~\cite{Sakakibara2023}. 
Instead, the fluorinated bilayer retains the canonical, cuprate-like fermiology. 
Because the residual self-doping can be tuned continuously by chemical pressure and epitaxial strain, these external control parameters provide a practical route to optimize the carrier balance while preserving the underlying low-energy electronic structure, much as proposed for the original infinite-layer nickelates~\cite{Held2024,Held2025}. However, the resulting changes in the Ni-$d_{x^2-y^2}$ filling remain modest, suggesting that these parameters act as fine-tuning rather than strong-doping mechanisms.

The magnetic properties reveal a second distinctive aspect of La$_3$Ni$_2$O$_5$F. The weak interlayer exchange produces an extensive near-degeneracy among competing collinear magnetic configurations that survives both structural relaxation and the inclusion of on-site Coulomb interactions. Compressive strain further suppresses the already small energy differences between these states while leaving the local Ni moments largely unchanged. Together, these results indicate a robust quasi-two-dimensional frustrated magnetic landscape, consistent with recent studies of aliovalently substituted infinite-layer nickelates~\cite{day-roberts_electron_2026} and pressurized single-layer Ruddlesden-Popper nickelates~\cite{de_vaulx_pressure_2025}, as well as with the absence of clear long-range magnetic order in the available experiments~\cite{Wernert2026}.

In this respect, La$_3$Ni$_2$O$_5$F exhibits an interesting resemblance to bulk FeSe~\cite{Glasbrenner2015}. Although the microscopic exchange mechanisms are different, both systems lie close to a manifold of nearly degenerate magnetic states whose relative stability is highly sensitive to lattice tuning. In FeSe, such magnetic frustration has been proposed as an important ingredient in its unconventional superconductivity. The analogous magnetic landscape identified here suggests that fluorinated nickelates may provide a complementary setting in which to explore the interplay between lattice degrees of freedom, magnetic frustration, and correlated electronic phases.

Finally, our results significantly extend the recent work of Song \textit{et al.}~\cite{Song2026a,Song2026b}, who attributed the self-doping in La$_3$Ni$_2$O$_5$F to an interstitial transmutation of rare-earth-derived states and identified a shallow magnetic-energy landscape from constrained-moment calculations of the FM configuration. Our calculations confirm that this shallow landscape extends beyond the FM channel and the itinerant picture, with multiple competing magnetic configurations forming an extended manifold of nearly degenerate states. Moreover, we show that both the rare-earth-derived self-doping and the magnetic competition can be continuously tuned by chemical pressure and epitaxial strain. Together with the emergence of a strain-controlled incipient charge instability, these results establish lattice engineering as a practical route for tailoring the electronic and magnetic properties of fluorinated nickelates.

\section{Conclusions}

Using first-principles calculations, we have characterized the electronic and magnetic properties of the newly synthesized double infinite-layer oxyfluoride La$_3$Ni$_2$O$_5$F. 
We confirm that the reported structure is dynamically stable based on its phonon spectrum, remaining stable also under epitaxial strain.

As the bilayer extension of the T' oxyfluoride La$_2$NiO$_3$F, La$_3$Ni$_2$O$_5$F exhibits a strongly cuprate-like electronic structure. In particular, it hosts a highly two-dimensional, weakly hybridized Ni-$d_{x^2-y^2}$ Fermi surface with a moderate self-doping contribution intermediate between the single-layer T' oxyfluoride and infinite-layer nickelate limits. These electronic features are remarkably robust against chemical pressure and epitaxial strain, which nevertheless provide controlled means to fine-tune the carrier balance.

At the same time, La$_3$Ni$_2$O$_5$F exhibits a strong tendency toward magnetism while retaining an extended manifold of nearly degenerate magnetic configurations. This near-degeneracy can be further enhanced by epitaxial strain, pointing to a frustrated quasi-two-dimensional magnetic landscape. Such competing low-energy states may suppress conventional long-range order and promote enhanced magnetic fluctuations, which could play an important role in the emergent low-energy physics of this system.

These results establish fluorinated bilayer nickelates, and more broadly the $R_{2}R'_{n-1}$Ni$_n$O$_{2n+1}$F family, as a promising platform for investigating emergent correlated phases in novel quantum materials.

\section*{Acknowledgements}
We acknowledge P. Toulemonde, V. Olevano, and M. Braun for stimulating and helpful discussions. We also acknowledge HPC resources from GRICAD and GENCI Grants 2022-AD010913948 and 2025-AD010916740, and the ANR-25-CE30-1574 for funding.

\bibliographystyle{apsrev4-2}
\bibliography{bib}

\clearpage
\onecolumngrid
\newpage
\renewcommand{\thefigure}{S\arabic{figure}}
\renewcommand{\thetable}{S\arabic{table}}
\setcounter{page}{1}
\setcounter{figure}{0}
\setcounter{table}{0}

\begin{center}
{\bf \Large Supplemental Material}\\[1em]

{\bf \large
Structural stability, electronic structure, and magnetism of the 
\\
$d^9$ double infinite-layer La$_3$Ni$_2$O$_5$F under chemical pressure and epitaxial strain
}\\[1em]

K. Madani, Q. N. Meier, and A. Cano\\[1ex]

{\it \small Univ. Grenoble Alpes, CNRS, Grenoble INP, Institut N\'eel, 25 Rue des Martyrs, 38042, Grenoble, France}

(\small \today)
\end{center}

\vspace{2em}

\begin{figure*}[h]
    \centering
    \includegraphics[width=0.7\linewidth]{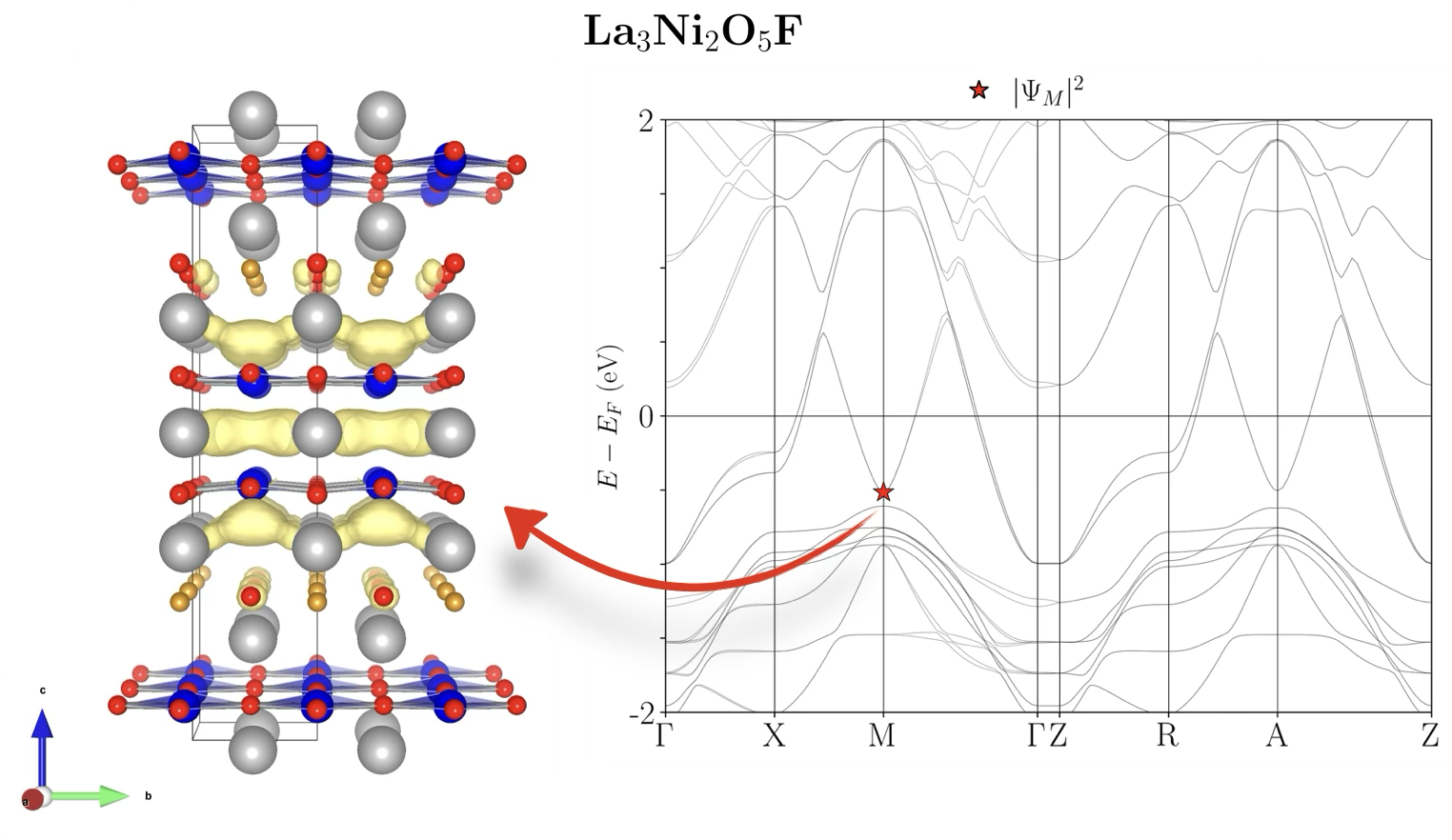}
    \caption{Iso-surface of $|\Psi_{nk}|^2$ with $n=$ self-doping band and $k=$ M-point.}
    \label{fig:Psi_M}
\end{figure*}

\begin{figure*}[h]
\centering
\includegraphics[width=1\linewidth]{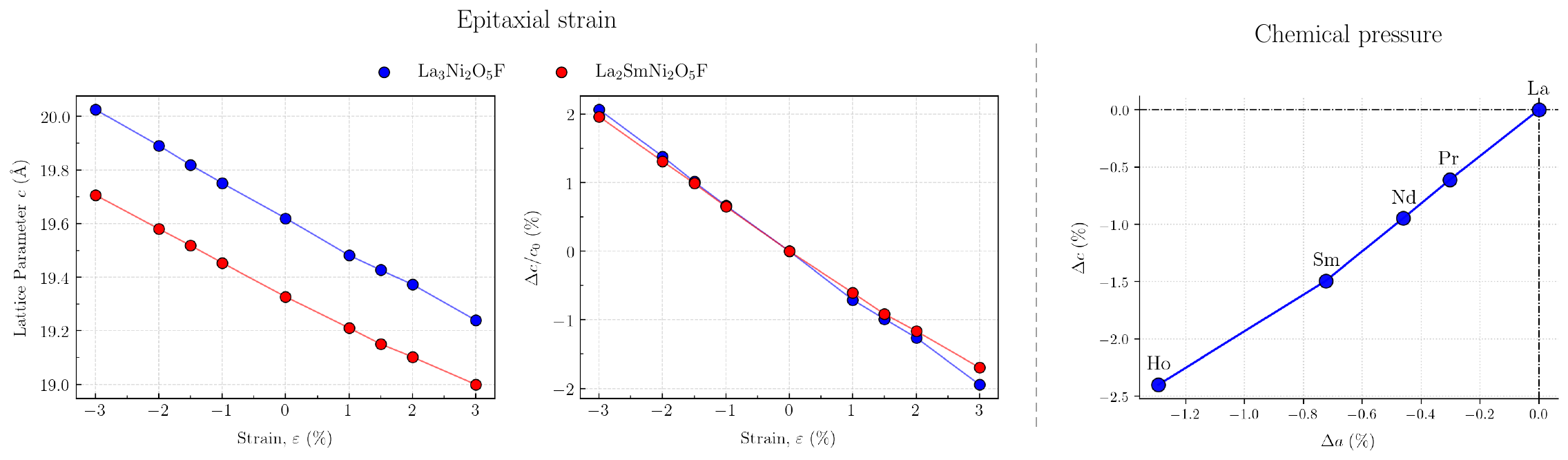}
\caption{Lattice parameters as a function of epitaxial strain $\varepsilon$ calculated for La$_3$Ni$_2$O$_5$F and La$_2$SmNi$_2$O$_5$F, and lattice-parameter change as a function of the La~$\to R$ substitution in La$_{2}R$Ni$_2$O$_{5}$F.
}
\label{fig:lattice_params}
\end{figure*}

\begin{figure*}[h]
\centering
\includegraphics[width=0.6\linewidth]{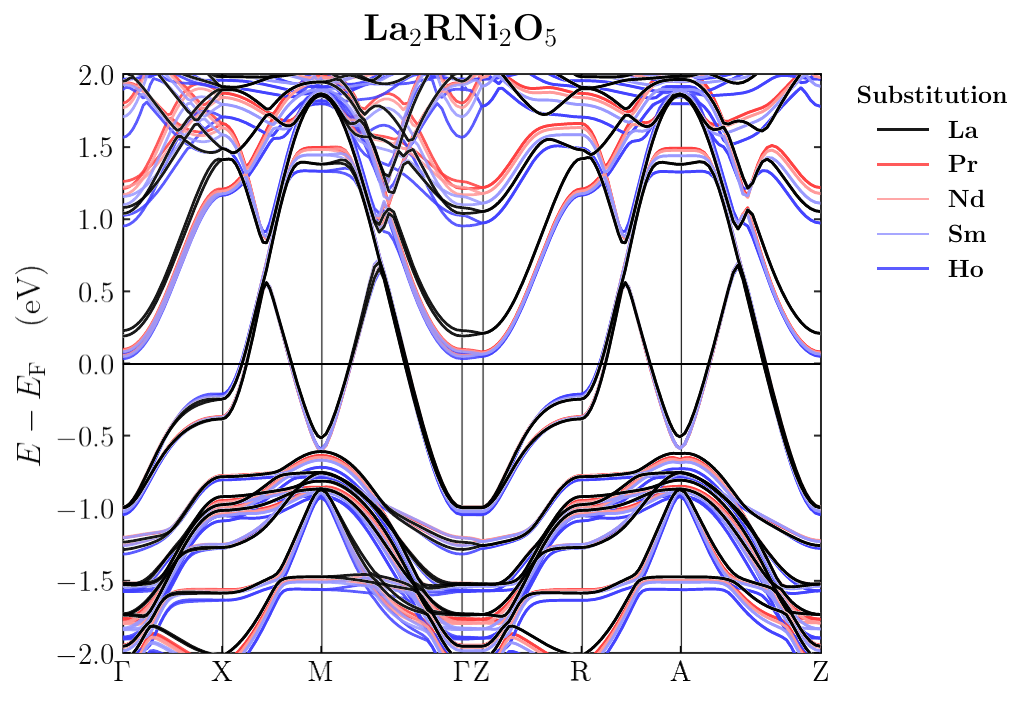}
\caption{Calculated band structures for the different rare-earth substitutions.}
\label{fig:bands_Re}
\end{figure*}

\begin{figure*}[h]
\centering
\includegraphics[width=\linewidth]{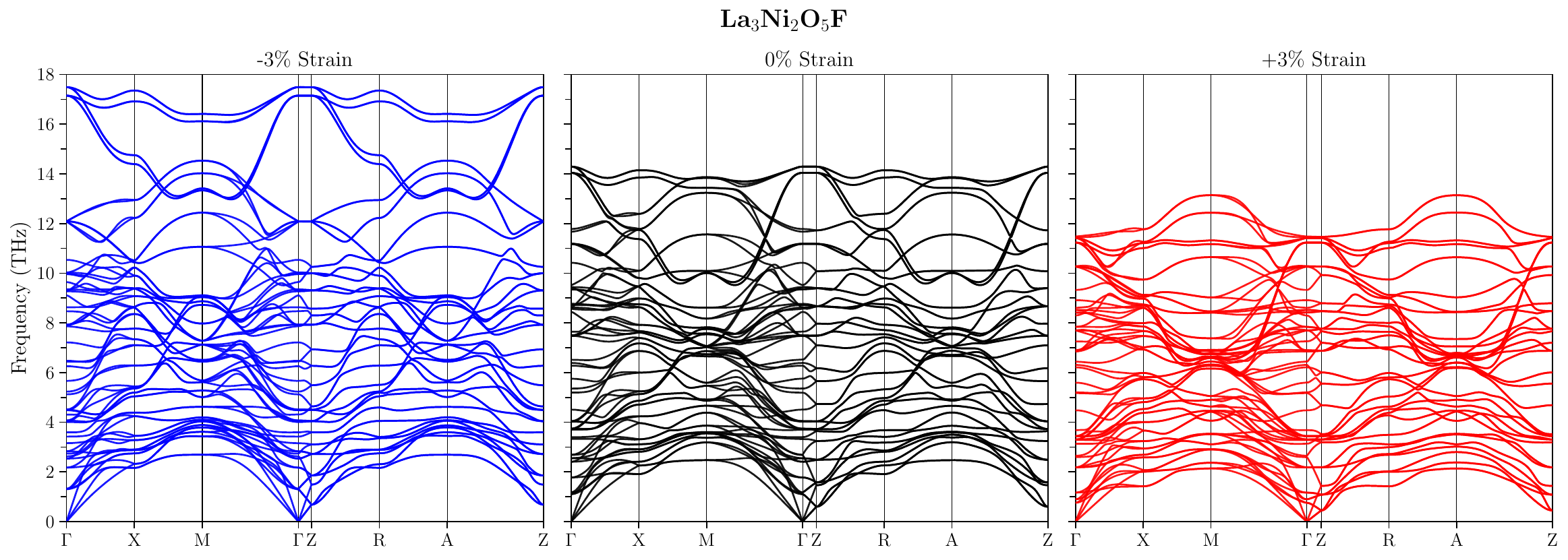}
\caption{Calculated phonon spectra of La$_3$Ni$_2$O$_5$F under varying epitaxial strain values ($\varepsilon = -3\%, 0\%,$ and $+3\%$) in the $2(\sqrt{2}\times\sqrt{2})\times1$ supercell .}
\label{fig:phonons-app}
\end{figure*}

\begin{figure*}[h]
\centering
\includegraphics[width=0.75\linewidth]{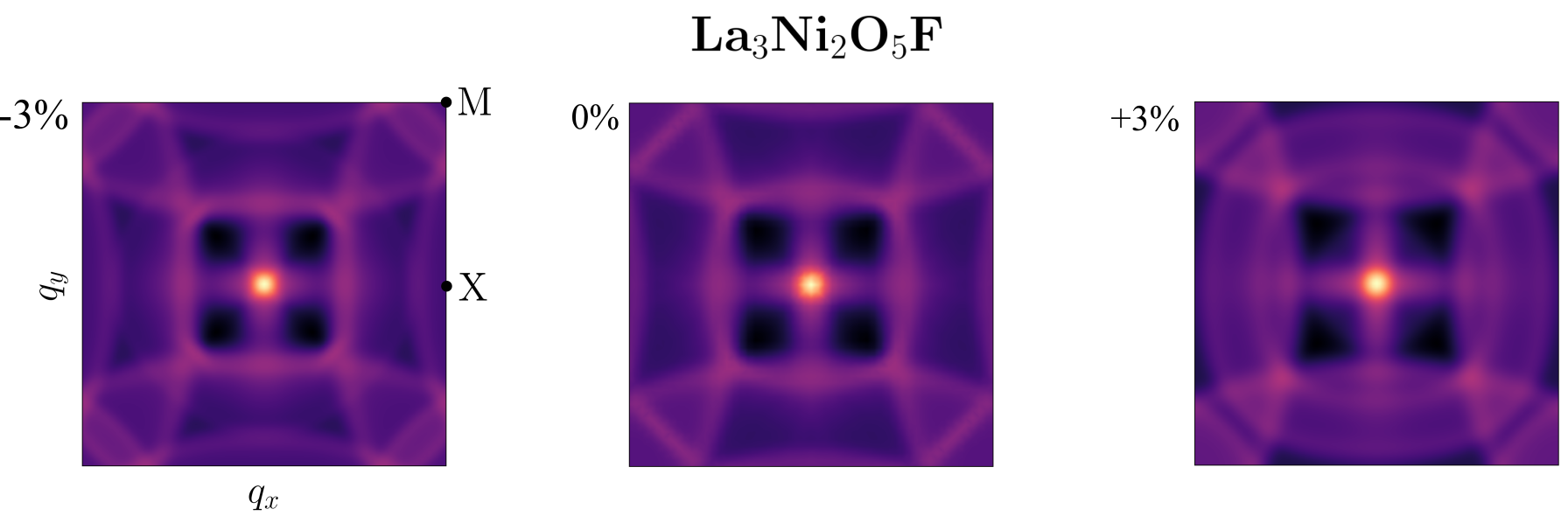}
\caption{
Calculated nesting function
$\underset{\omega\to 0}{\lim}
\chi''_0(\mathbf{q},\omega)/\omega = \sum_{\mathbf{k},n,m}
\delta(\varepsilon_{n,\mathbf{k}}-E_F)\,
\delta(\varepsilon_{m,\mathbf{k}+\mathbf{q}}-E_F)$
in the $q_z=0$ plane for La$_3$Ni$_2$O$_5$F at
$\varepsilon=-3\%$, $0\%$, and $+3\%$ epitaxial strain, with
high-symmetry points $M$ and $X$ indicated. The central peak at
$\mathbf{q}=0$ is the trivial self-nesting contribution
($\mathbf{k}=\mathbf{k}+\mathbf{q}$ for every point on the Fermi
surface).}
\label{fig:dd-app}
\end{figure*}

\begin{figure*}[h]
    \centering
    \includegraphics[width=1\linewidth]{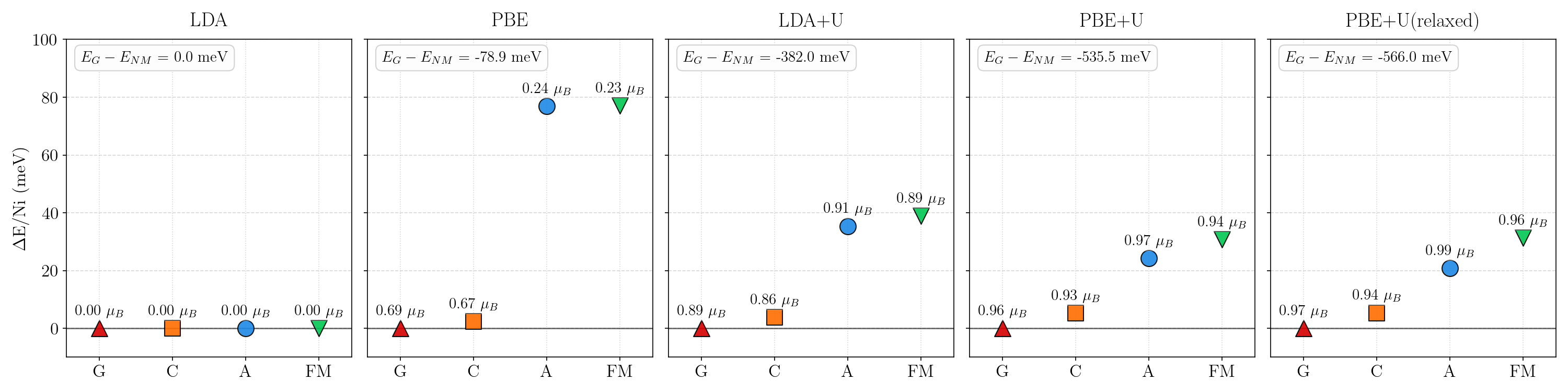}
    \caption{Relative energy per Ni atom with respect to the G-AFM configuration for $\text{La}_3\text{Ni}_2\text{O}_5\text{F}$, calculated using LDA, LDA+$U$, GGA, and GGA+$U$ approximations, as well as GGA+$U$ with spin-polarized structural relaxation. The magnetic stabilization energy relative to the non-magnetic (NM) state ($\Delta E_{\text{G-AFM} - \text{NM}}$) is explicitly annotated for each case.}
    \label{fig:mag_LDA_GGA_appendix}
\end{figure*}

\begin{figure*}[h]
\centering
\begin{minipage}{0.48\linewidth}
\centering
\includegraphics[width=\linewidth]{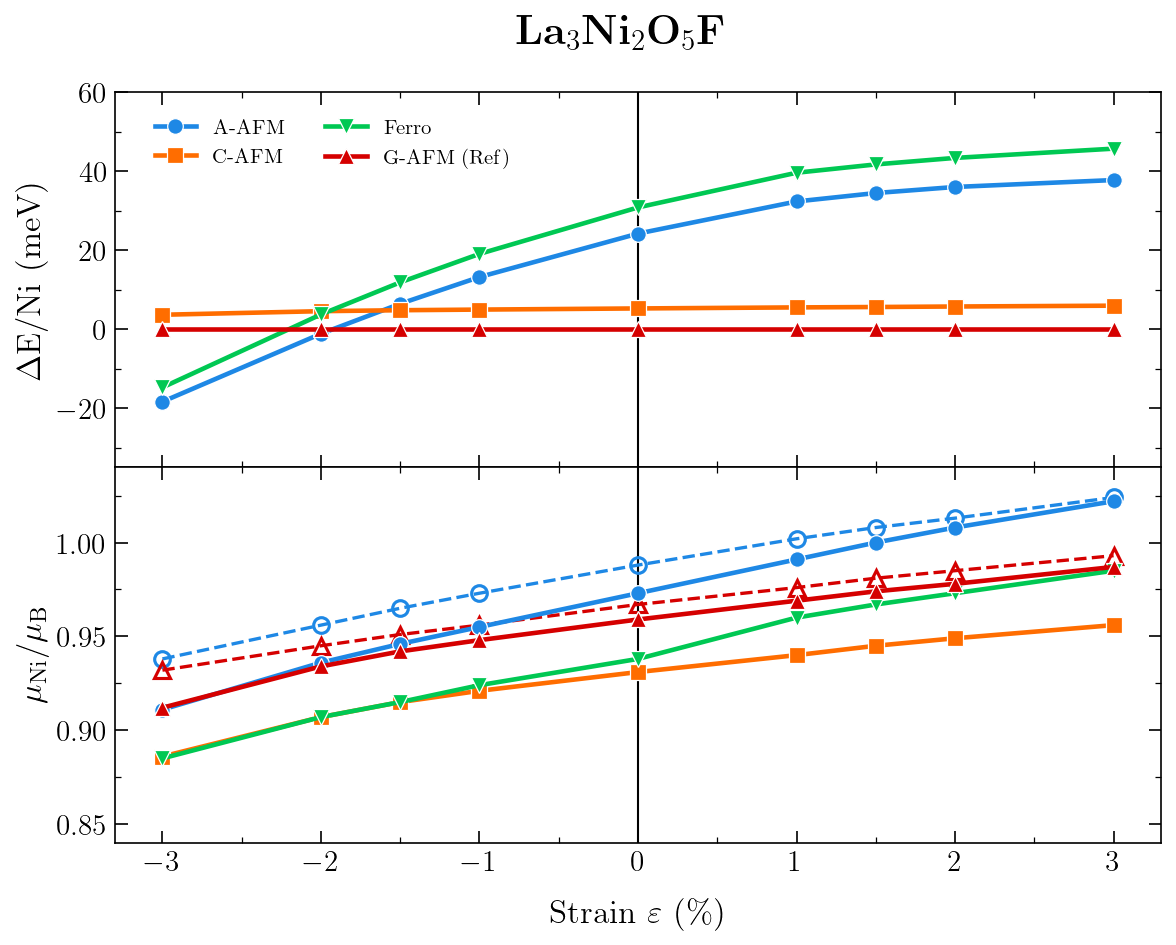}
\end{minipage}
\hfill
\begin{minipage}{0.48\linewidth}
\centering
\includegraphics[width=\linewidth]{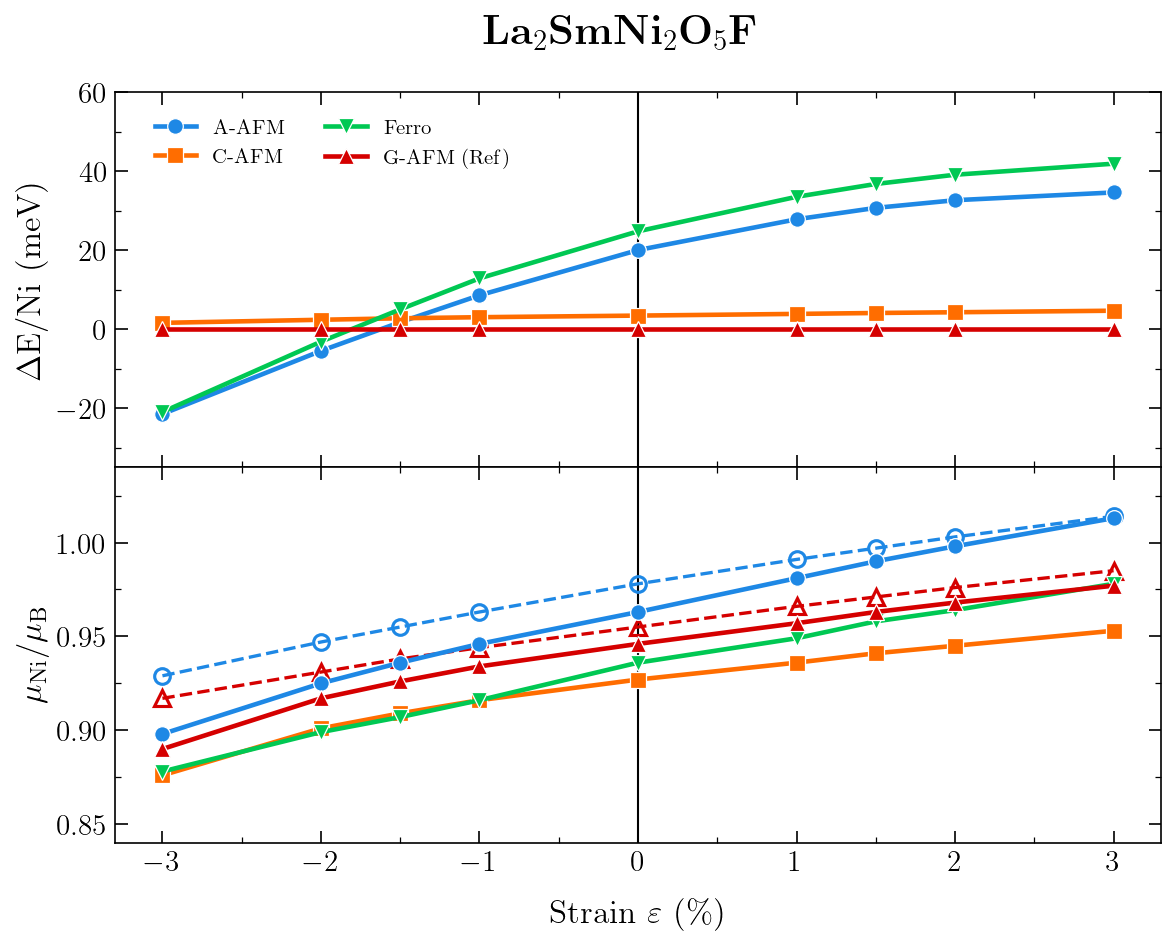}
\end{minipage}
\caption{Relative energy per Ni atom with respect to the G-AFM configuration for La$_3$Ni$_2$O$_5$F (left) and La$_2$SmNi$_2$O$_5$F (right) as a function of strain, and corresponding Ni magnetic moments.}
\label{fig:magnetism_appendix}
\end{figure*}

\end{document}